\documentclass[twocolumn,pra,showpacs,floatfix,amsmath,amsfonts]{revtex4-1}
\usepackage[latin9]{inputenc}
\usepackage{color}
\usepackage{amsmath}
\usepackage{amssymb}
\usepackage{graphicx}
\usepackage{esint}

\makeatletter
\@ifundefined{textcolor}{}
{%
 \definecolor{BLACK}{gray}{0}
 \definecolor{WHITE}{gray}{1}
 \definecolor{RED}{rgb}{1,0,0}
 \definecolor{GREEN}{rgb}{0,1,0}
 \definecolor{BLUE}{rgb}{0,0,1}
 \definecolor{CYAN}{cmyk}{1,0,0,0}
 \definecolor{MAGENTA}{cmyk}{0,1,0,0}
 \definecolor{YELLOW}{cmyk}{0,0,1,0}
}

\usepackage{color}
\usepackage{graphics}
\usepackage{epsfig}
\newcommand{\be}{\begin{equation}}
\newcommand{\ee}{\end{equation}}
\newcommand{\bea}{\begin{eqnarray}}
\newcommand{\eea}{\end{eqnarray}}
\newcommand{\bes}{\begin{subequations}}
\newcommand{\ees}{\end{subequations}}
\newcommand{\PT}{\mathcal{PT}}

\newcommand{\p}{{\cal P}}
\newcommand{\T}{{\cal T}}

\newcommand{\hA}{\hat{A}}

\newcommand{\bM}{\mathbf{M}}

\newcommand{\bb}{\mathbf{b}}

\newcommand{\bk}{\mathbf{k}}
\newcommand{\bq}{\mathbf{q}}

\newcommand{\br}{\mathbf{r}}

\newcommand{\rt}{{\rm t}}
\newcommand{\rr}{{\rm r}}

\newcommand{\erez}{\varepsilon_{\rm res}}

\newcommand{\rev}[1]{\textcolor{black}{#1}}
\newcommand{\new}[1]{\textcolor{black}{#1}}

\allowdisplaybreaks

\makeatother

\begin{document}

\title{Broadband \rev{quasi}-$\PT$ Symmetry Sustained by Inhomogeneous Broadening of the Spectral Line
}

\author{ D. M. Tsvetkov$^1$, V. A. Bushuev$^1$, V. V. Konotop$^{2}$, B. I. Mantsyzov$^1$  }

\affiliation{ $^1$Department of Physics, M. V. Lomonosov Moscow State University, 	Moscow 119991, Russia
	\\
 $^{2}$Departamento de F\'{i}sica and Centro de F\'{i}sica Te\'orica e Computacional, Faculdade de Ci\^encias, Universidade
de Lisboa, Campo Grande, Ed. C8, Lisboa 1749-016, Portugal
}

\date{\today}
\begin{abstract}
 It is shown that inhomogeneous broadening of the spectral line of active impurities may sustain  simultaneously parity ($\p$) and time ($\T)$ symmetries of a medium, in a finite range of field frequencies, what  is  forbidden by the causality principle in media without broadening. If a spectral width of a propagating pulse is less than the inhomogeneous broadening, the medium for such a pulse becomes \rev{quasi-}$\PT$-symmetric. The effect of the {broadband} \rev{quasi-}$\PT$ symmetry in a finite frequency domain, is illustrated on examples of unidirectional diffraction of pulses in Bragg and Laue geometries, propagating in photonic crystals.
\end{abstract}


\maketitle

\section{Introduction}
{Parity-time ($\PT$) symmetry of a non-Hermitian operator, whose spectrum can be purely real~\cite{BenBoet}, from the practical point of view is a highly demanding requirement. 
In optical applications one-dimensional $\PT$ symmetry requires exact balance between gain and loss described by the imaginary part of the dielectric permittivity $\varepsilon(\xi)$, where $\xi$ is one of spatial coordinates depending on the statement of the problem. Such balance condition can be loosely interpreted as $\varepsilon(\xi)= \varepsilon^*(-\xi)$~\cite{Muga,Christod2007,Christod2008}. The formulated relation however is written, and thus is applicable, only for a given frequency of light, $\omega$.  The fundamental importance of this last constraint, stems from the causality principle, usually} expressed through the Kramers-Kronnig relations. Indeed, the latter can be satisfied only for isolated frequencies~\cite{Vinogradov}. That is why, the major development of the $\PT$-symmetric optics is achieved, so far, within the paraxial approximation proposed in \cite{Christod2008}, rather than with optical pulses, as initially suggested in~\cite{Muga}. The only examples of optical pulses in $\PT$-symmetric systems are considered in models of quasi-monochromatic wave-packets propagating in coupled active and lossy nonlinear waveguides when the material dispersion can be neglected. Such models are described by  (nonlinear) Schr\"odinger equations with gain and losses~\cite{coupled_NLS} (see also~\cite{KYZ} and references therein).  
 
\rev{Among other applications, optical $\PT-$symmetric systems can be considered as tools for manipulating light. Thus, to make this possibility of light control universal, it is desirable to have it applicable to as broader types of pulse as possible. In particular, it is known that in many applications in spectroscopy,   optical communications,  and  nonlinear nonlinear optics, as shorter a pulse is as more efficient is its usage. Therefore it is natural to consider a possibility of applying the ideas of $\PT$ symmetry to as short optical pulses as possible. Presenting of the solutions of this problem is the goal of this paper.}

 $\PT$ symmetry is also a very delicate property: even small imperfectness of a system may destroy its guiding properties leading to instabilities,  i.e. may result in  complex eigenvalues in the spectrum. In particular, breaking of a $\PT$-symmetric phase in a discrete chain with random impurities~\cite{Kottos}, and  unbounded growth of squared field characteristics of a randomly perturbed $\PT$-symmetric coupler~\cite{ZK-random} have been reported.

Based on these arguments one could arrive at a natural conclusion that the $\PT$ symmetry is not an appropriate framework for dealing with short optical pulses, especially when the gain and losses do not have perfectly matched spatial landscapes. In this paper we show that this conclusion is not always true, and quite counter-intuitively, namely imperfectness of atomic resonances may allow for sustaining {\em broadband \rev{quasi}}-$\PT$ symmetry in strongly dispersive media. \rev{Although such systems are not exactly $\PT$-symmetric, the average dispersion of the complex dielectric permittivity is suppressed strongly enough, making the medium effectively $\PT$-symmetric even for short optical pulses.} Thus such media are suitable for  exploring pulse dynamics in $\PT$-symmetric environment. As examples, we describe propagation of picosecond pulses in $\PT$-symmetric photonic crystals under the condition of dynamical Bragg diffraction in the Bragg and Laue geometries. 

Propagation of monochromatic beams in $\PT$-symmetric media is asymmetric. As examples we mention unidirectional invisibility of Bragg grating~\cite{Lin} and directed transport of a $\PT$-coupler~\cite{Ramezani2010},  
non-reciprocity of Bloch oscillations in $\PT$-symmetric crystals~\cite{Longhi}, incident-angle dependence of mode amplification in waveguides with $\PT$-symmetric core profiles~\cite{giant}, asymmetry of  waves diffracted and reflected by a periodic array in the Laue scheme of Bragg diffraction~\cite{KonMan,Bushuev2017}, etc. All these phenomena were considered for monochromatic waves at frequencies at which $\PT$ symmetry of the guiding medium is verified. In this paper we report  asymmetry of  propagation of {\em short {picosecond} pulses} in $\PT$-symmetric layered media with dispersion.

The paper is organized as follows. In Sec.~\ref{sec:broad} we describe how inhomogeneous broadening allows to create quasi-$\PT$ symmetry for a finite domain of frequencies (while  the exact $\PT$ symmetry can be valid for a single frequency). General aspects of the diffraction in a quasi-$\PT$-symmetric medium are described in Sec.~\ref{sec:general}. In sections~\ref{sec:Bragg} and \ref{sec:Laue} we consider examples of the diffraction in Bragg and in Laue geometries, respectively. The results as summarized in Conclusion. Some technical details of calculations are given in Appendixes.

\section{Effect of inhomogeneous broadening} 
\label{sec:broad}

To describe the physics of the phenomenon we consider a photonic crystal (the host medium) with real dielectric permittivity, $\varepsilon_{\rm hm}(\xi)$, having period $d$ in $\xi$-direction (below $\xi$ will be either $x$ or $z$), i.e., $\varepsilon_{\rm hm}(\xi+d)=\varepsilon_{\rm hm}(\xi)$.  The host medium  is doped by active impurities, which are two-level atoms, resulting in gain and absorption described by the permittivity component,  $\varepsilon_{\rm res} (\xi,\omega)$. Since active media are highly dispersive we account for the frequency dependence of  $\varepsilon_{\rm res} (\xi,\omega)$, while the dispersion of the host medium is neglected [$\varepsilon_{\rm hm}(\xi)$ does not depend on $\omega$]. We assume that spatial distribution of inversion of the excited atoms is described by a periodic odd function $w(\xi)$ with zero average and having the period of the host medium: $w(\xi)=-w(-\xi)=w(\xi+d)$. 

Let the impurity atoms have resonant frequencies $\omega_0'$ which deviate from a central frequency $\omega_0$. The distribution of these deviations $g(\omega_0'-\omega_0)$  determines the resonant component of the dielectric permittivity, $\erez (\xi,\omega)$. The {two-level quantum oscillator} model (see e.g.~\cite{Allen,model}) allows one to represent $\erez (\xi,\omega)=-iw(\xi)\tilde{\varepsilon}(\omega)$, where
\begin{eqnarray}
\label{epsilon}
 \tilde{\varepsilon}(\omega)= i\beta  \int_{-\infty}^{\infty}\frac{g(\Delta-\Delta_0)}{\Delta+i/T_2}d\Delta,
\end{eqnarray}
$\beta=4\pi N\mu^2/\hbar$, $N$ is the concentration of resonant atoms, $\mu$ is the dipole moment,  $T_2$ is the homogeneous transverse relaxation time of a single atom, $\Delta=\omega-\omega_0'$ and by $\Delta_0=\omega-\omega_0$
(we use the commonly accepted notations~\cite{Allen}). 

We assume that the time of the inhomogeneous transverse relaxation, $T_2^*$, which characterizes the distribution $g(\Delta-\Delta_0)$,  is much less than the relaxation time of a single atom: $T_2^*\ll T_2$. Then, the deviations of the resonant frequencies $\omega_0'$ from the central frequency $\omega_0$ are constrained by $|\omega_0'-\omega_0|\lesssim 1/T_2^*$. Furthermore, we assume that the detuning of the field frequency from $\omega_0$ is much smaller than the line width of the inhomogeneous broadening, i.e.,  $|\Delta_0|\ll 1/T_2^*$.
 
Now we show that, subject to the imposed conditions on the resonant medium, the odd distribution of the  inversion $w(\xi)$ assures, with high accuracy, the $\PT$ symmetry of the dielectric permittivity for pulses of a finite duration $\tau$ satisfying $\tau\gg T_2\gg T_2^*$.  Indeed, the resonant component of the dielectric permittivity (\ref{epsilon}) can be expressed as a sum ($j=0,1$)
 \begin{equation}
 \label{G}
 \tilde{\varepsilon}
 = \beta 
 \left(g_0+ig_1\right),\,\,\, g_{j}= (T_2)^{j-1}\!\int_{-\infty}^{\infty}\frac{g(\Delta-\Delta_0)\Delta^j  }{\Delta^2+1/T_2^2}d\Delta.
 \end{equation}

 For the assumed odd distribution of the inversion $w(\xi)$, the terms involving $g_0$ and $g_1$ represent $\PT$-symmetric and anti-$\PT$-symmetric parts of the dielectric permittivity $\erez(\xi,\omega)$, respectively. Hence,  for the model to obey $\PT$ symmetry, one has to require $|g_1|\ll |g_0|$. 
 To study the conditions when it happens we explore the Doppler-Maxwellian distribution~\cite{model}
 \begin{eqnarray}
 \label{Doppler}
 g(\Delta-\Delta_0)=\frac{T_2^*}{\sqrt{2\pi}}e^{-(\Delta-\Delta_0)^2(T_2^*)^2/2},
 \end{eqnarray}
 which is the most used one and in which  $1/T_2^*$ is the standard deviation from the central detuning $\Delta_0$. In the limit of zero inhomogeneous broadening $1/T_2^*\to 0 $, when $g(\Delta-\Delta_0)\to \delta(\Delta-\Delta_0)$,  we obtain that the medium is $\PT$-symmetric for a monochromatic wave at the central frequency $\omega_0$  ($g_1=0$ at $\Delta=\Delta_0$) and does not possess any symmetry for all other frequencies from the spectrum of a pulse-spectrum. 
 
Now we show that presence of sufficiently large inhomogeneous broadening, $1/T_2^*\gg 0$, sustains $\PT$ symmetry for a  {\em finite} spectral domain centered at $\omega_0$. Indeed, using expressions (\ref{G}) and (\ref{Doppler}) we compute in the leading order of the limit ${T_2^*}/{T_2},\, T_2^*\Delta_0\ll 1$ the estimates   (see Appendix~\ref{app:estimates})
\begin{eqnarray}
\label{estimates}
g_0 \sim T_2^*, \quad g_1 \sim   (T_2^*)^2\Delta_0
\end{eqnarray}
 Thus we indeed obtain the required condition $g_1/g_0\sim T_2^*\Delta_0 \ll1$ allowing for neglecting the anti-$\PT$ symmetric part of the dielectric permittivity. We emphasize, that the above estimates  are not restricted to model (\ref{Doppler}). The essential requirement used in their derivation [see (\ref{A3}) and (\ref{A6})] is the sufficiently fast (Gaussian in our case) decay of the distribution $g(\Delta )$ at $\Delta\to \infty$. Thus, any distribution of the type $g(\Delta )f(\Delta )$ where $g(\Delta )$ is given by (\ref{Doppler}) and $f(\Delta )$ is any function decaying at $\Delta\to \infty$, will lead to (\ref{estimates}).

{The described system can be experimentally realized by doping a periodic Bragg grating (see e.g. \cite{Bragg_technol}) or a layered medium of transparent materials, by resonant ions like Nd$^{3+}$, Er$^{3+}$, or Ti$^{3+}$, which are characterized by sufficiently large inhomogeneous line broadening, typically of about  $10^{13}\,$Hz \cite{ions}. The periodic inversion of excited ions having the required symmetry, can be achieved, for example, by means of a periodic mask whose symmetry axis is shifted by a quarter of the period of the host material \rev{\cite{Lupu}}.}
	
Let us assume that the spectral width of the incident pulse, $\delta\omega=2/\tau$, coincides with the  resonant frequency $\omega_0$. Then $\Delta_0\lesssim 2/\tau$ and the relation between the anti-$\PT$-symmetric and $\PT$-symmetric parts is rewritten as $g_1/g_0\sim T_2^*/\tau \ll1$. Thus, the photonic crystal is an effectively $\PT$-symmetric medium for pulses with of the duration $\tau\gg T_2^*$ in a similarity with a monochromatic wave which corresponds the infinite duration. The above conditions can be achieved in solid-state media. For instance,  all the requirements are satisfied for picosecond pulses, $\tau> 10^{-12}$s in neodymium glass where $T_2^*\lesssim 10^{-13}$s and $T_2\sim10^{-12}$s~\cite{Herrmann}. 

\begin{figure}[h]
	\includegraphics[width=\columnwidth]{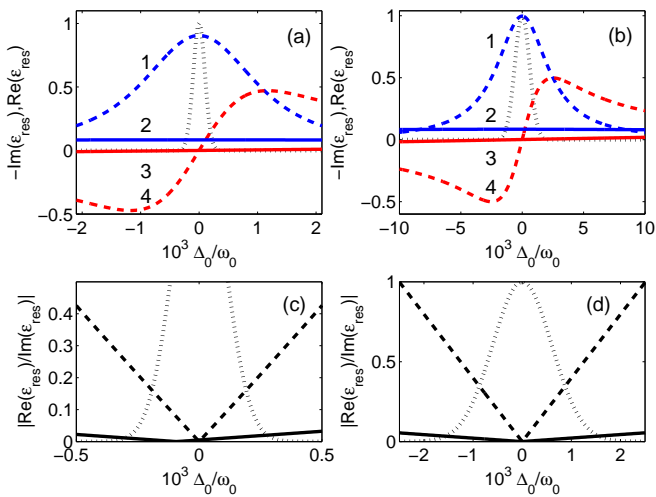}
	\caption{{Upper panels:  $-$Im$(\erez)$ (blue lines 1 and 2) and  Re$(\erez)$ (red lines 3 and 4) versus frequency detuning at two different values inhomogeneous broadening plotted for the point where $w(\xi)=1$ for the parameter choice $\beta=1/T_2$.  In (a)  $\gamma_2^*=0.0005$ (dashed lines)  and  $\gamma_2^*=0.02$ (solid lines) compared with the spectrum of 6 ps pulse (dotted line) for $\gamma_2=0.002$. In (b)  $\gamma_2^*=0.0002$ (dashed lines)  and  $\gamma_2^*=0.05$ (solid lines) compared with the spectrum of 1 ps pulse (dotted line) for $\gamma_2=0.005$. Lower panels: in (c) and (d) dependences of $|\mbox{Re}(\erez)/\mbox{Im}(\erez)|$ on the frequency detuning corresponding to panels (a) and (b). The left and right panels correspond to dynamics shown below in Fig.~\ref{fig:three} and Fig.~\ref{fig:four}.} 
}
	\label{fig:one}
\end{figure} 
In Fig.~\ref{fig:one} we plot {the real and imaginary parts of the resonant dielectric permittivity for two different choices of the dimensionless parameters $\gamma_2=2/(\omega_0 T_2)$ and $\gamma_2^*=2/(\omega_0 T_2^*)$, compared with the spectra of two pico-second pulses (dotted lines). One observes the transition from the strong frequency dependent $\varepsilon_{\rm res}$  at small $\gamma_2^*$ (dashed lines) to the suppressed frequency dependence for relatively large $\gamma_2^*$ (solid lines).  To characterize quantitatively the deviation of the profile of the dielectric permittivity from the $\PT$-symmetric one, in lower panels we show the frequency dependence of the parameter 
	$|g_1/g_0|=|\mbox{Re}(\erez)/\mbox{Im}(\erez)|$. The observed improvement of the $\PT$-symmetry is larger than an order of magnitude (zero  value of this parameter correspond to the ideal frequency independent $\PT$ symmetry).}

 \section{Pulse diffraction in a \rev{quasi-}$\PT$-symmetric medium}
\label{sec:general}
 
\begin{figure}[h]
 	\includegraphics[width=\columnwidth]{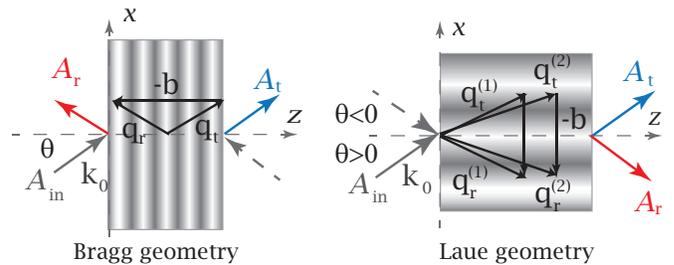}
 	\caption{ Schematic illustration of the Bragg (left panel) and Laue (right panel) diffractions. Solid and dashed arrows show the central wave-vectors of the pulses incident from the left and right (Bragg diffraction) and with positive and negative incidence angles (Laue diffraction). Red and blue arrows correspond to the pulses reflected  and transmitted by the photonic cristal. The central wavevectors of resonantly interacting waves in the photonic crystal, $\bq_\rt$ and $\bq_\rr$ are also shown. In the Laue geometry there are two pairs of such vectors.}
 	\label{fig:two}
 \end{figure}

  In order to illustrate effects of the $\PT$ symmetry in a finite frequency domain,  we note that in an experiment the easiest parameter to manipulate is the inhomogeneous broadening $\gamma_2^*$. Therefore we consider the effect of $\gamma_2^*$ on unidirectional diffraction of pico-second pulses incident on a periodic medium. Let an $s$-polarized pulse 
  \begin{equation}
  \label{eq:Einit}
    E_{\rm in}=A_{\rm in}(\br,t)\exp\left(i\bk_0\br-i\omega_0 t\right)  
  \end{equation}
  with the central frequency $\omega_0$ and the wavevector $\bk_0$, $k_0=\omega_0/c=2\pi/\lambda_0$, $c$ being the light speed in vacuum, and with slowly varying amplitude $A_{\rm in} (\br,t)$, is applied to a photonic crystal. We explore the diffraction in the Bragg geometry (the left panel of Fig.~\ref{fig:two}) and in the Laue geometry (the right panel of Fig.~\ref{fig:two}). In both cases the propagation direction is $z$, while the crystal boundary is along $x$. Let the dielectric perimittivity of the host medium is 
  periodically modulated: $\varepsilon_{\rm hm}=\varepsilon_0+\varepsilon'\cos(b\xi)$ where $b=2\pi/d$  is the modulus of the reciprocal lattice vector, $\bb$,  $\varepsilon'$ is the depth of periodic  modulation, and hereafter $\xi=z$ in the Bragg geometry and $\xi=x$ in the Laue geometry.
  We also assume that the inversion distribution has the form $-w(\xi)=\sin(b \xi)$. Thus, inside the crystal, the field $E(\br,\omega)$ is governed by the Helmholtz equation
 \begin{eqnarray}
 \label{Helmholtz}
 \nabla^2 E+k^2[\varepsilon_0+\varepsilon' \cos(b \xi)+i\tilde{\varepsilon} (\omega) \sin(b \xi) ]E=0,
 \end{eqnarray}
 where $k=\omega/c$.

We address the simplest nontrivial diffraction patterns of resonantly interacting waves,  
in Bragg~\cite{Authier} and in Laue~\cite{Skorynin} geometries, simultaneously.  For shallow grating
\begin{eqnarray}
\label{shallow}
|\varepsilon'|,|\tilde{\varepsilon}|, \ll\varepsilon_0
\end{eqnarray}
 the resonance conditions are met if the incidence angle $\theta$ is close to the Bragg angle, i.e. $\theta_B=\arcsin(\varepsilon_0-b^2/4k_0^2)^{1/2}$ for the Bragg geometry and  $\theta_B=\arcsin(b/2k_0)$ for the Laue geometry.  Different aspects of Bragg scattering of monochromatic waves under normal incidence was studied in earlier publications~\cite{Lin,Bragg}. The $\PT$-symmetric Laue diffraction of monochromatic waves was considered in~\cite{KonMan,Bushuev2017,Brandao}.

\new{In the two mode approximation~\cite{Skorynin} (see also Appendixes~\ref{app:Bragg} and \ref{app:Laue})} the field in the photonic crystal is searched in the form  
\begin{eqnarray}
\label{eq:inPC}
E(\br,t)=\left[A_{\rm t}(\br,t)e^{i\bq_\rt\br}+A_{\rm r}(\br,t)e^{i\bq_\rr\br}\right]e^{-i\omega_0t}
\end{eqnarray}
  where $\br=(x,z)$, $\bq_\rt$ is the central wavevector  of the wave transmitted by the first boundary of the crystal, $A_{\rm t}(\br,t)$ is a slowly varying envelope of the trasmitted pulse, while $A_{\rm r}(\br,t)$  is an envelope of either reflected (in Bragg geometry) or diffractionally reflected (in Laue geometry) pulses. The  relation between wavevectors of the resonant modes, $s\bb=\bq_\rt-\bq_\rr$, where $s=+1$ ($s=-1$) for left (right) incidence in the Bragg geometry or for positive (negative) angle of incidence in the Laue geometry, depends on the case: $\bb=(0,0,b)$  and  $\bb=(b, 0,0)$ for the Bragg and  Laue geometries, respectively (see Fig.~\ref{fig:two}).
 
 Subject to the above  assumptions, the Fourier transforms of the pulse amplitudes $\hat{A}_{\rm t,r}$, combined in a vector-column $\hA=(\hA_{\rm t},\hA_{\rm r})^T$, at a given frequency $\omega$ solve the system (see Appendixes~\ref{app:Bragg} and~\ref{app:Laue} for details):  
\begin{eqnarray}
\label{E}
 \bM\hA=q_\rt^2\hA, \quad \bM=\frac{k^2}{2}\left(\begin{array}{cc}
2\varepsilon_0 & \varepsilon'+s\tilde{\varepsilon}(\omega) 
\\ 
\varepsilon'-s\tilde{\varepsilon}(\omega)  & 2\varepsilon_0 + 2\alpha 
\end{array}\right),
\end{eqnarray}
   where we introduce the small factor $\alpha=\bb(2s\bq_\rt-\bb)/k^2$, $|\alpha|\ll 1$ describing deviation from the exact Bragg condition $\bb(2s\bq_\rt-\bb)=0$. Now the conventional dispersion relation, expressing the link among $\omega$, $q_{\rt x}$ and $q_{\rt z}$, and obtained from the secular equation 
   \begin{eqnarray}
   \label{dispersion}
   \mbox{det}(\bM-q_\rt^2)=0
   \end{eqnarray}
    can be recast in the form $q_\rt=q_\rt(\omega, \alpha)$. Then for any pair of the fixed parameters $\omega$, $\alpha$, Eq.~(\ref{E})  is transformed to the standard eigenvalue problem for the matrix $\bM$ with $q_\rt^2$ being the eigenvalue. For generic choice of the parameters in the plane $(\omega,\alpha)$ the matrix $\bM$ is diagonalizable. If however, for a given frequency, the amplitude of the gain and losses is chosen to satisfy   $\varepsilon'= \pm \tilde{\varepsilon}(\omega)$, and simultaneously the exact Bragg condition, $\alpha=0$ is fulfilled, then $\bM$ becomes defective and the exceptional point (EP) emerges. This can be viewed as   generalization of the EP for the well studied normal incidence in the Bragg geometry where $\bq_\rt=s\bb/2$ (i.e. $\alpha=0$).
 Thus, the occurrence of EP depends on both distribution of active impurities  and on the geometry of incidence.  
 
\new{For calculations of the pulse dynamics one can use the spectral method ~\cite{Skorynin}. It consists of the following  steps: the field of the incident pulse is represented in a form of the Fourier integral; the Bragg diffraction problem is solved for a monochromatic plane wave; the Fourier synthesis is used to obtain the field at each point of the photonic crystal at a given time. Two examples are considered in the next sections.}

\section{Bragg geometry} 
\label{sec:Bragg}

\new{Starting with the Bragg geometry we define 
$$\bk_0=({k}_{0x},{k}_{0z})={k}_{0}(\sin \theta , s{{k}_{0}}\cos \theta), $$
where $s=1$ ($s=-1$) in the case of a pulse is incident on the left (right) surface of the photonic crystal. 
The incident wave packet (\ref{eq:Einit}) now has the form  
\begin{eqnarray}
\label{eq:incid_Bragg_E}
E_{\text{in}}(x,t)=A_{\text{in}}(x,t)\exp(ik_{0x}x-i\omega _{0}t). 
\end{eqnarray}
where $A_{\text{in}}(x,t)$ is a slowly varying amplitude.
}

 Let $\hat{A}_{\text{in,t,r}}(\Omega) $ be  the frequency spectra of the incident ("in"),  transmitted ("t"), and diffractionally reflected ("r") waves, \rev{where $\Omega =\omega -{{\omega }_{0}}$}. Taking into account the continuity of the tangential components of the wave vectors ${{q}_{\text{t}x}}={{k}_{0x}}+\Omega \sin \theta /c$ on the photonic crystal surfaces $z=0$ (if $s=1$) and $z=l$ (if $s=-1$), we have 

\begin{eqnarray}
\label{B1}
A_{\text{in}}(x,t)&=& \int_{-\infty }^{\infty }{{\hat{A}}_{\text{in}}}(\Omega ){e^{-i\Omega (t-x\sin \theta /c)}}d\Omega,   
\\
\label{B81}
A_{\text{t,r}}(\mathbf{r},t)&=&\int_{-\infty }^{\infty }\hat{A}_{\text{t,r}}(\Omega)e^{-i\Omega (t-{{\tau }_{\text{t}\text{,r}}})}d\Omega ,       
\end{eqnarray}
where ${{\tau }_{\text{t}\text{,r}}}=(x\sin \theta \pm sz\cos \theta )/c$. 

\new{
Standard algebra, outlined in Appendix~\ref{app:Bragg} yields
\begin{eqnarray}
\label{B19}
\hat{A}_{\text{t}}(\Omega )&=&\frac{{{{\hat{A}}}_{\text{in}}}(\Omega )}{1-P}\left( 1-\frac{{{R}_{1}}}{{{R}_{2}}} \right)e^{isq_{\text{t}z}^{(1)}\rev{l}},      
\\
\label{B20}
\hat{A}_{\text{r}}(\Omega )&=&\frac{{{R}_{1}}{{{\hat{A}}}_{\text{in}}}(\Omega )}{1-P}\left[ 1-e^{is(q_{\text{t}z}^{(1)}-q_{\text{t}z}^{(2)})l} \right].                                        
\end{eqnarray}
where 
\begin{eqnarray}
\label{B12}
q_{\text{t}z}^{(1,2)}=\frac{sb}{2}\mp \frac{{{k}^{2}}}{2b}\sqrt{{{\alpha }^{2}}-[{{{{\varepsilon }'}}^{2}}-{{{\tilde{\varepsilon }}}^{2}}(\omega )]},  
\end{eqnarray}
is obtained from the dispersion relation (\ref{dispersion}), and
\begin{eqnarray}
\label{B15}
P=\frac{{{R}_{1}}}{{{R}_{2}}}e^{is(q_{\text{t}z}^{(1)}-q_{\text{t}z}^{(2)})l}.           
\end{eqnarray}
}

\begin{figure}[h]
	\includegraphics[width=\columnwidth]{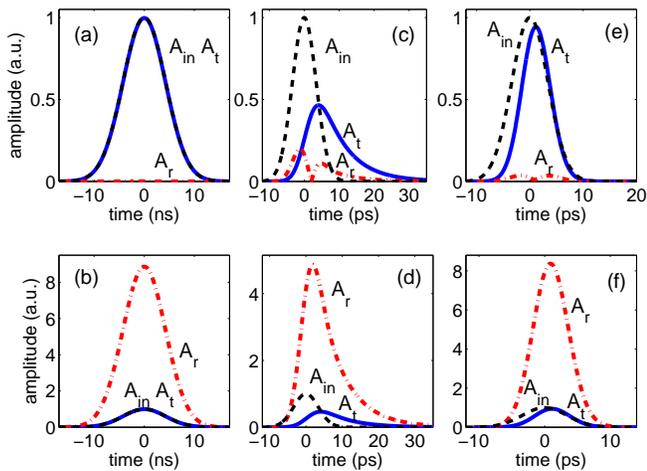}
 
	\caption{ Temporal profiles of the transmitted (blue solid lines) and reflected (red dash-dot lines) pulses generated by the incident Gaussian pulse (dashed line) upon left (upper row) and right (lower row) incidence. The inhomogeneous broadening is $\gamma_2^*=0.0005$ in {(a)-(d)} and  $\gamma_2^*=0.02$ in (e),(f).  {In (a), (b) the duration of the incident pulse is $6\,$ns. Note that in (a), (b) the input and output signals  are indistinguishable on the scale of the figure, while in (a) the reflected wave is indistinguishable from zero.} In  (c)--(f) the incident pulse duration $\tau=6\,$ps, i.e. $\delta\omega/\omega_0=0.0001$, $\lambda_0=0.8\,\mu$m and the Bragg angle of incidence is $\theta=63.4^0$, while the photonic crystal characteristics are $d=0.5657\,\mu$m, $\varepsilon_0=1.3$, $\varepsilon'=0.008$, Re$[\tilde{\varepsilon}(\omega_0)]/\varepsilon'=0.9999$, and
		$\gamma_2=0.002$. The total width of the crystal is $0.2\,$mm. Note the different axis scales in different panels.}
	\label{fig:three}
\end{figure}

In Fig.~\ref{fig:three} we show the amplitudes $\left| {A}_{\text{t}\text{,r}}(t) \right|$  illustrating the phenomenon of unidirectional  diffraction reflection of an incident Gaussian pulse (\ref{eq:incid_Bragg_E})  with
\begin{eqnarray}
\label{eq:incid_Bragg_A}
 A_{\text{in}}(x,t)=A\exp \left[-\frac{{(ct-x\sin\theta)}^2}{c^2\tau ^{2}}\right],
\end{eqnarray}
where $\tau$ is the pulse duration,
by a layered medium in the Bragg geometry for the left (upper row) and right (lower row) incidence.   \rev{The phenomenon consists in propagation of the pulse incident from the left hand side of the layered medium [Figs.~\ref{fig:three} (a) and (e)] with negligible reflection and strongly enhanced diffractionally reflected signal in the case of right incidence [Figs.~\ref{fig:three} (b) and (f)].}  

The characteristics of the medium are chosen as in Fig.~\ref{fig:one} (a), (c). The pulse is chosen such that in the absence of the active impurities, $\tilde{\varepsilon}=0$, its frequency spectrum  almost entirely belongs to the forbidden gap, and thus the pulse is totally reflected. It is also assumed that the distribution of active impurities and the angle of incidence correspond to the EP. In  presence of weak inhomogeneous broadening, a wide quasi-monochromatic nano-second pulse is totally transmitted [reflected] at the left, Fig.~\ref{fig:three} (a) [right Fig.~\ref{fig:three} (b)] incidence. However for the same medium but for a short pico-second pulse having spectral width $\delta\omega/\omega_0$ comparable with $\gamma_2^*\sim\delta\omega/\omega_0\ll\gamma_2$, the $\PT$-symmetric effects are strongly suppressed. Now one observes partial transmission and reflection of the pulse for both left [Fig.~\ref{fig:three} (c), c.f. panel (a)] and right [Fig.~\ref{fig:three} (d), c.f. panel (b)] incidence. Transmitted (blue solid lines) and reflected (red dash-dot solid lines)  pulses emerge strongly deformed. Significant increase of the inhomogeneous broadening, ensuring $\gamma_2^*\gg\gamma_2\gg\delta\omega/\omega_0$ results in enhancement of the unidirectional reflection: there is almost total transmission of the pulse for the left [Fig.~\ref{fig:three} (e)] and enhanced reflection for the right [Fig.~\ref{fig:three} (f)] incidence. In this last case, the Gaussian shape of the incident pulse is almost restored. \rev{It is to be mentioned that the observed enhancement is not surprising, because energy is not conserved, and the eigenmodes of a non-Hermitian operator are not orthogonal, i.e. they experience energy exchange. That is why, the intensity of a superposition of modes in a $\PT$-symmetric medium while remaining bounded can be increased several orders of magnitudes~\cite{SchKon}.  }   The transmitted pulse almost coincides with incident one [Fig.~\ref{fig:three} (e)], while the reflected signal is strongly enhanced [Fig.~\ref{fig:three} (f)]. \rev{Finally, we mention that increasing of the pulse duration, say passing from picosecond to nano-second pulses, under the conditions described in Fig.~\ref{fig:three} (e) and Fig.~\ref{fig:three} (f), will make the non-$\PT$-symmetric effects even weaker, and hence will display qualitatively the same evolution as the picosecond pulses show.} 

\section{Laue geometry}
\label{sec:Laue}

\new{In the  Laue scheme of Bragg diffraction (right panel of Fig.~\ref{fig:two}), we use spatio-temporal Fourier transform. Now the incident field at $z=0$, given by (\ref{eq:Einit}) has the slowly varying amplitude, which we express trough the spectral amplitude ${{\hat{A}}_{\text{in}}}(K,\Omega )$ 
\begin{eqnarray}
\label{C2}
 A_{\rm in}(x,t)=\iint\limits_{-\infty }^{\infty }\hat{A}_{\text{in}}(K,\Omega )e^{iKx-i\Omega t}dKd\Omega. 
\end{eqnarray}
Respectively, the transmitted and reflected spectral amplitudes are introduced through the amplitudes $A_{\rm t,r}(K,\Omega )$ defined by (\ref{eq:inPC}):
\begin{eqnarray}
\label{C2}
 A_{\rm g}(\br,t)=\iint\limits_{-\infty }^{\infty }{\hat{A}_{\text{g}}}(K,\Omega )e^{iKx+i{q_{\text{g}z}}z-i\Omega t}dKd\Omega ,
\end{eqnarray}
where g=t,r and} ${{q}_{\text{t}z}}={{q}_{\text{r}z}}$ is to be determined from the dispersion relation. 
The direct algebra gives (see Appendix~\ref{app:Laue})
\begin{equation}
\label{Ar}
{{\hat{A}}_{\text{g}}}(K,\Omega )={{\hat{A}}_{\text{g1}}}{{e}^{iq_{\text{t}z}^{(1)}z}}+{{\hat{A}}_{\text{g2}}}{{e}^{iq_{\text{t}z}^{(2)}z}},
\end{equation}
where ${{\hat{A}}_{\text{r}j}}={{R}_{j}}{{\hat{A}}_{\text{t}j}}$ ($j=1,2$), 
\begin{equation}
 \label{C9}
{{\hat{A}}_{\text{t}1}}=-\frac{(1+{R}_{S}){R}_{2}}{{R}_{12}}{\hat{A}}_{\text{in}}, 
\quad {\hat{A}}_{\text{t}2}=\frac{(1+{R}_{S}){R}_{1}}{{R}_{12}}{\hat{A}}_{\text{in}}.   
\end{equation}
 Here ${{R}_{12}}={R}_{1}-{R}_{2}$, ${{R}_{S}}=({{k}_{z}}-{{f}_{S}})/({{k}_{z}}+{{f}_{S}})$ is the diffraction-modified Fresnel reflection coefficient, and ${{f}_{S}}=(q_{\text{t}z}^{(2)}{{R}_{1}}-q_{\text{t}z}^{(1)}{{R}_{2}})/{{R}_{12}}$.

\begin{figure}[h]
 	\includegraphics[width=1.0\columnwidth]{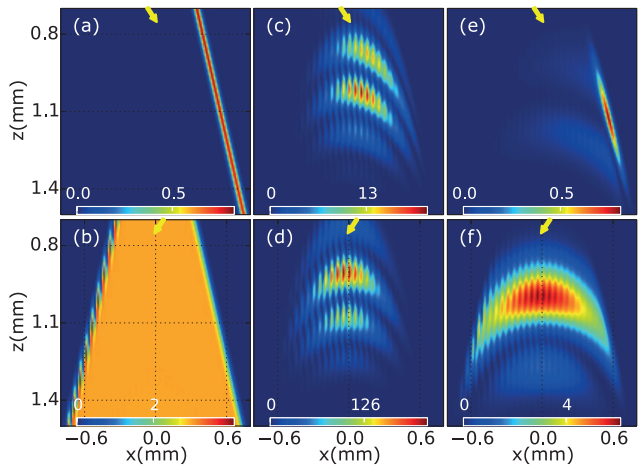}
 	\caption{Snapshots of the intensity \rev{(in arbitrary units)} distribution at $t=5\,$ps of a {quasi-monochromatic pulse having the input duration $\tau=1\,$ns} in (a), (b) and of a {short} $\tau=1\,$ps pulse in (c)--(f), with positive (upper row) and negative (lower row) Bragg angles of incidence (yellow arrows on the top of each panel). Inhomogeneous relaxation $\gamma_2^*=0.0002$ in (a)-(d) and $\gamma_2^*=0.05$ in (e),(f). Other parameters are   $\lambda_0=0.8\,\mu$m, $\varepsilon_0=1.3$, $\varepsilon'=0.008$,  Re$[\tilde{\varepsilon}(\omega_0)]/\varepsilon'=0.9999$, $d=0.8\,\mu$m, $\gamma_2=0.005$, $r_0=30\,\mu$m, \rev{$\theta=30^{\circ}$.}  }
 	\label{fig:four}
 \end{figure}

Strongly asymmetric diffraction of a pulse through a layered structure in the Laue geometry can be obtained at exactly Bragg angle of incidence. {Now the characteristics of the medium are chosen as in Fig.~\ref{fig:one} (b), (d).} We consider an incident pulse 
\begin{eqnarray}
\label{eq:incid_Laue}
A_{\rm in}(x,t)=A\exp\left[-\frac{(x\cos\theta)^2}{r_0^2}- \frac{(ct-x\sin\theta)^2}{c^2\tau^2}\right]
\end{eqnarray}
where $r_0$ and $\tau$ are its radius and duration, at the surface $z=0$. 
 In Fig.~\ref{fig:four} (a), (b) we show the snapshots of the intensity distributions of the above pulse {for 1 ns duration, i.e. for a quasi-monochromatic wave for which the medium can be viewed as approximately $\PT$-symmetric. We observe the expectable strong asymmetry of propagation. If however, the duration of the pulse is reduced to 1 ps, such that $\gamma_2^*\ll\delta\omega/\omega_0<\gamma_2$, the $\PT$-symmetry effects become strongly suppressed [panels (c) and (d)]. Now the  shapes of the pulses with positive and negative angles of incidence weakly differ from each other. Strong asymmetry, is recovered (i.e. $\PT$-symmetry is restored) for the picosecond pulse at} sufficiently large inhomogeneous broadening, $\gamma_2^*\gg\gamma_2>\delta\omega/\omega_0$ [Fig.~\ref{fig:four} (e), (f)].  \rev{Like in the case of Bragg geometry, the pulse propagation shown in the last two panels will be closely reproduced if the duration is increasing, say to nanoseconds, because narrowing spectral width of the incident pulse brings the system even closer to the exact $\PT$ symmetry.}
 
\begin{figure}[h]
	\includegraphics[width=\columnwidth]{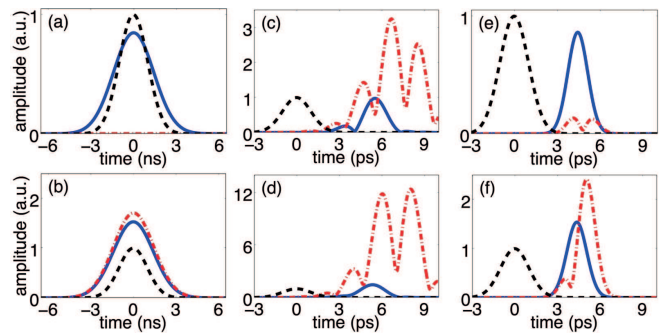}
	\caption{\rev{The time dependencies of the modules of the  field amplitudes at the propagation distance $z=1\ \text{mm}$ in the points $x=0$ [red dash-dot line], $x=0.495\ \text{mm}$ [blue lines in (c), (e)], $x=-0.495\ \text{mm}$ [blue lines in (d), (f)], for the positive (top panels) and negative (bottom panels) angles of incidence. The incident Gaussian pulse at $z=0$ is shown by dashed lines. The pulse duration is: $\tau =1\ \text{ns}$ in (a), (b) and $\tau =1\ \text{ps}$ in (c)-(f).  In (a)-(d) $\gamma _{2}^{*}=0.0002$ and in (e),(f) $\gamma _{2}^{*}=0.05$. The other parameters are as in Fig.~\ref{fig:four}}.
	}
	\label{fig:five}
\end{figure}

 \rev{Figure~\ref{fig:four} shows spatial distributions of the fields in the case of diffraction in the Laue geometry at a given moment in time. In order to make the fact of the existence of broadband quasi-$\PT$ symmetry more evident in this case, in Fig.~\ref{fig:five} we also present   the temporal profiles of the field amplitudes modules in different points of photonic crystal at the depth $z=1\,$mm. The red dash-dot curves correspond to the point $x=0$, the blue solid lines to the point $x=0.495$mm (the positive angle of incidence; top panels) and to $x=-0.495\text{mm}$ (the negative angle of incidence; bottom panels). The incident Gaussian pulse at $z=0$ is shown by dashed line. In the case of a weak inhomogeneous broadening line $\gamma _{2}^{*}=0.0002$ [Fig.~\ref{fig:five} (a)-(d)], the curves in Fig.~\ref{fig:five} (a), (b) for a quasi-monochromatic nanosecond pulse strongly differ in shape and amplitude from the respective curves showin in Fig.~\ref{fig:five} (c), (d) for a picosecond pulse. In the last case the amplitudes of the fields increase considerably and large oscillations appear at the point $x=0$. Obviously, quasi-$\PT$ symmetry is broken. When the inhomogeneous broadening is increased to a value $\gamma _{2}^{*}=0.05$, the pulses become much closer in amplitude and shape to those in the case of quasi-$\PT$ symmetry [cf. Fig.~\ref{fig:five} (e), (f) and Fig.~\ref{fig:five} (a), (b)]. Thus, we see a significant restoration of $\PT$ symmetry, or quasi-$\PT$ symmetry. Further increase in the parameter $\gamma_{2}^{*}>0.05$ leads to an even better coincidence of the fields in the $\PT$-symmetric and quasi-$\PT$-symmetric photonic crystals.}


\section{Conclusion}  

To conclude we have shown that inclusion of inhomogeneous relaxation of active impurities allows for implementation of \rev{broadband quasi-}$\PT$ symmetry hold in a finite frequency range, rather than for a single frequency. For any pulse characterized by the spectral width smaller than the inhomogeneous line broadening, such a medium is close to $\PT$-symmetric. As illustration of how \rev{broadband quasi-}$\PT$ symmetry can affect pulse propagation we describe unidirectional diffraction reflection by a photonic crystal in Bragg and Laue geometries. For a given pulse duration such diffraction strongly depends of the inhomogeneous broadening.

The reported results allow direct extension in at least two directions. First, as soon as the spectral characteristics are involved, the approach paves the way for a frequency control on the  pulse dynamics in the $\PT$-symmetric media.   Second,  by simple change of the host medium modulation from even to odd and simultaneous change of the inversion from odd to even, the created medium becomes anti-$\PT$ symmetric. Such media recently received particular attention in theoretical~\cite{anti-PT-theor} and experimental studies~\cite{anti-PT-exper}. Thus the  effect of the inhomogeneous broadening allows for creating anti-$\PT$-symmetric media for finite frequency domains.

\acknowledgments
The work of D.M.T., V.A.B., and B.I.M. was supported by the Russian Foundation for Basic Research, grant 18-02-00556-a. 

\renewcommand{\theequation}{\Alph{section}\arabic{equation}}

\appendix

\section{$\PT$ and anti-$\PT$ symmetric parts of the dielectric permittivity}
\label{app:estimates}

Using the Doppler-Maxwellian distribution (\ref{Doppler})
we compute the following set of transformations: 
\begin{eqnarray}
\label{A2}
\frac{\sqrt{2\pi }}{T_{2}^{*}} g_1
= \int_{-\infty }^{\infty }{\frac{{{e}^{-{{(\Delta -{{\Delta }_{0}})}^{2}}{{({{T}^{*}})}^{2}}/2}}\Delta d\Delta }{{{\Delta }^{2}}+1/T_{2}^{2}}}
\nonumber  \\
= \int_{-\infty }^{\infty }\frac{{{e}^{-{{(\nu -{{\nu }_{0}})}^{2}}}}\nu d\nu }{{{\nu }^{2}}+{{\alpha }^{2}}} 
%
= e^{-\nu _{0}^{2}} {F}_{1}(\nu_{0},\alpha ),
\end{eqnarray}
where
\begin{eqnarray}
{F}_{1}(\nu_{0},\alpha )=\int_{-\infty }^{\infty }{\frac{{{e}^{-{{\nu }^{2}}}}\sinh (2\nu {{\nu }_{0}})\nu d\nu }{{{\nu }^{2}}+{{\alpha }^{2}}}},
\end{eqnarray} 
$\displaystyle{ \alpha =T_{2}^{*}/\sqrt{2}T_{2}\ll 1}$, 
and $\displaystyle{\nu_{0}=T_2^{*}\Delta_{0}/\sqrt{2}}$.
 
Using the expansion of the hyperbolic sine in the Taylor series we have 
\begin{multline}
\label{A3}
F_{1}(\nu_{0},\alpha ) 
=\sum_{p=0}^{\infty }{\frac{1}{(2p+1)!}\int_{-\infty }^{\infty }{\frac{{{e}^{-{{\nu }^{2}}}}{{(2\nu {{\nu }_{0}})}^{2p+1}}\nu d\nu }{{{\nu }^{2}}+{{\alpha }^{2}}}}}
\\ 
=\sum_{p=0}^{\infty }{\frac{{{(2{{\nu }_{0}})}^{2p+1}}}{(2p+1)!}\int_{-\infty }^{\infty }{\frac{{{e}^{-{{\nu }^{2}}}}{{\nu }^{2(p+1)}}d\nu }{{{\nu }^{2}}+{{\alpha }^{2}}}}}
\\
\xrightarrow{\alpha \to 0}\sum\limits_{p=0}^{\infty }{\frac{{{(2{{\nu }_{0}})}^{2p+1}}}{(2p+1)!}\int_{-\infty }^{\infty }{{{e}^{-{{\nu }^{2}}}}{{\nu }^{2p}}d\nu }}= \\ 
=2\sum\limits_{p=0}^{\infty }{\frac{{{(2{{\nu }_{0}})}^{2p+1}}p!}{(2p+1)!}}=2\sqrt{\pi }{{\nu }_{0}}+{\mathrm O}(\nu _{0}^{2}).
\end{multline}
Thus, in the leading order we obtain 
\begin{eqnarray}
\label{A4}
{{g}_{1}}\sim{{(T_{2}^{*})}^{2}}{{\Delta }_{0}} \;\;\; \mathrm{at}\; \;\;  \frac{T_{2}^{*}}{{{T}_{2}}}, \; T_{2}^{*}{{\Delta }_{0}}\to 0.
\end{eqnarray}                                       
For the $\PT$-symmetric part of the resonant dielectric permittivity we calculate 
\begin{multline}
\label{A5}
 \frac{\sqrt{2\pi }}{T_{2}^{*}}{g}_{0}=\frac{1}{{T}_{2}} \int_{-\infty }^{\infty }{\frac{{{e}^{-{{(\Delta -{{\Delta }_{0}})}^{2}}{{({{T}^{*}})}^{2}}/2}}d\Delta }{{{\Delta }^{2}}+1/T_{2}^{2}}}=\\
= \frac{T_{2}^{*}}{\sqrt{2}T_2}\int_{-\infty }^{\infty }\frac{{{e}^{-{{(\nu -{{\nu }_{0}})}^{2}}}}d\nu }{{{\nu }^{2}}+{{\alpha }^{2}}}
= \frac{T_{2}^{*}}{\sqrt{2}T_2}e^{-\nu _{0}^{2}}{{F}_{0}}({{\nu }_{0}},\alpha ), 
\end{multline}
where 
\begin{multline}
\label{A6}
 F_{0}(\nu _{0},\alpha )=\int_{-\infty }^{\infty }{\frac{{{e}^{-{{\nu }^{2}}}}\cosh (2\nu {{\nu }_{0}})d\nu }{{{\nu }^{2}}+{{\alpha }^{2}}}=}
 \\
 =\sum_{p=0}^{\infty }{\frac{{{(2{{\nu }_{0}})}^{2p}}}{(2p)!}\int_{-\infty }^{\infty }{\frac{{{e}^{-{{\nu }^{2}}}}{{\nu }^{2p}}d\nu }{{{\nu }^{2}}+{{\alpha }^{2}}}}}= \\ 
 =\int_{-\infty }^{\infty }{\frac{{{e}^{-{{\nu }^{2}}}}d\nu }{{{\nu }^{2}}+{{\alpha }^{2}}}}+\sum\limits_{p=1}^{\infty }{\frac{{{(2{{\nu }_{0}})}^{2p}}}{(2p)!}\int_{-\infty }^{\infty }{\frac{{{e}^{-{{\nu }^{2}}}}{{\nu }^{2p}}d\nu }{{{\nu }^{2}}+{{\alpha }^{2}}}}}\\
 \xrightarrow{\alpha \to 0}\frac{\pi }{2\alpha }+\sum\limits_{p=1}^{\infty }{\frac{{{(2{{\nu }_{0}})}^{2p}}}{(2p)!}\int_{-\infty }^{\infty }{{{e}^{-{{\nu }^{2}}}}{{\nu }^{2(p-1)}}d\nu }} 
 \\ 
=\frac{\pi }{2\alpha }+2\sum\limits_{p=1}^{\infty }{\frac{{{(2{{\nu }_{0}})}^{2p}}}{{{4}^{p-1}}(2p-1)p!}}. 
\end{multline}
Thus 
\begin{eqnarray}
\label{A7}
{{g}_{0}}\sim\frac{\sqrt{\pi }}{2\sqrt{2}}T_{2}^{*} \; \;\; \mathrm{at}\; \;\; \frac{T_{2}^{*}}{{{T}_{2}}}, \; T_{2}^{*}{{\Delta }_{0}}\to 0. 
\end{eqnarray}

\section{Transmitted and reflected pulse amplitudes in the Bragg geometry} 
\label{app:Bragg}         
   
To find the $z$-components ${{q}_{\text{t}z}}$ and ${{q}_{\text{r}z}}$, as well as the relation between the Fourier amplitudes ${\hat{A}}_{\text{t,r}}$ we substitute (\ref{B1}) and (\ref{B81}) into Eq.~(\ref{Helmholtz}). Equating terms with equal exponents we obtain  
\begin{eqnarray}
\label{B9}
\begin{split}
({{\varepsilon }_{0}}-q_{\text{t}}^{2}/{{k}^{2}}){{{\hat{A}}}_{\text{t}}}+{{\varepsilon }_{-s}}{{{\hat{A}}}_{\text{r}}}=0, \\
 {{\varepsilon }_{s}}{{{\hat{A}}}_{\text{t}}}+[{{\varepsilon }_{0}}-{{({{\mathbf{q}}_{\text{t}}}-s\mathbf{b})}^{2}}/{{k}^{2}}]{{{\hat{A}}}_{\text{r}}}=0, \end{split}
\end{eqnarray}
 [which is system (\ref{E}) in the Bragg geometry], where
\begin{eqnarray}
\label{epsilon2}
{\varepsilon}_{1}=\left[{\varepsilon}'-\tilde{\varepsilon}\left(\omega\right)\right]/2,
\quad {\varepsilon}_{-1}=\left[{\varepsilon}'+\tilde{\varepsilon}\left(\omega\right)\right]/2.
\end{eqnarray}
 Thus the amplitudes of the transmitted, ${{\hat{A}}_{\text{t}}}(\Omega )$, and diffracted, ${{\hat{A}}_{\text{r}}}(\Omega )$, waves are related as   ${\hat{A}_{\text{r}}}=R{\hat{A}_{\text{t}}}$, where 
\begin{eqnarray}
\label{B10}
R=-\left(Q_{z}^{2}-q_{\text{t}z}^{2}\right)/{{\varepsilon }_{-s}}{{k}^{2}},            
\end{eqnarray}
${{Q}_{z}}=sQ$, $Q={{({{\varepsilon }_{0}}{{k}^{2}}-q_{\text{t}x}^{2})}^{1/2}}$ is the modulus of $z$-component of the wavevector for a homogeneous medium with the dielectric constant ${{\varepsilon }_{0}}$.  

In limit (\ref{shallow}) we represent ${{q}_{\text{t}z}}$ in the form ${{q}_{\text{t}z}}={{Q}_{z}}+{{\tilde{q}}_{\text{t}z}}$, where $|{{\tilde{q}}_{\text{t}z}}|\ll Q$ and approximate $q_{\text{t}z}^{2}\approx Q_{z}^{2}+2{{Q}_{z}}{{\tilde{q}}_{\text{t}z}}$. Then the solvability of Eqs. (\ref{B9}) takes the form 
\begin{eqnarray}
\label{B11}
4Q(b-Q)\tilde{q}_{\text{t}z}^{2}+2s{{k}^{2}}Q\alpha {{\tilde{q}}_{\text{t}z}}+{{\varepsilon }_{1}}{{\varepsilon }_{-1}}{{k}^{4}}=0,         \end{eqnarray}
where $\alpha =b(2Q-b)/{{k}^{2}}$ characterizes the detuning of incident angle $\theta$ from the exact Bragg condition. Near the exact Bragg condition, the solution of Eq. (\ref{B11}) takes the form of (\ref{B12}), 
while the partial reflection coefficients (\ref{B10}) have the following form 
\begin{eqnarray}
\label{B13}
{{R}_{1,2}}=-\frac{\alpha \pm s\sqrt{{{\alpha }^{2}}-[{{{{\varepsilon }'}}^{2}}-{{{\tilde{\varepsilon }}}^{2}}(\omega )]}}{2{{\varepsilon }_{-s}}},   
\end{eqnarray}
The amplitudes of the transmitted waves in the photonic crystal ${{\hat{A}}_{\text{t1}\text{,t2}}}$   are found from the boundary conditions on the left surface $z=0$  
(i.e. at ${{k}_{0z}}>0$ and $s=1$ for the pulse incident from the left) 
\begin{eqnarray}
\label{B14.a}
\begin{split}
{{\hat{A}}_{\text{t}1}}+{{\hat{A}}_{\text{t}2}}&={{\hat{A}}_{\text{in}}}(\Omega ),
\\
{{R}_{1}}{{\hat{A}}_{\text{t}1}}\exp (iq_{\text{t}z}^{(1)}l)
+&{{R}_{2}}{{\hat{A}}_{\text{t}2}}\exp (iq_{\text{t}z}^{(2)}l)=0. 
\end{split}
\end{eqnarray}
From these equations we obtain $(j=1,2)$
\begin{eqnarray}
\label{B15a}
{{\hat{A}}_{\text{t}j}}=\frac{{{(-P)}^{j-1}}}{1-P}{{\hat{A}}_{\text{in}}}(\Omega ),
 \end{eqnarray}
where $P$ is defined in (\ref{B15}).

For a pulse incident from the right on the surface $z=l$ (${{k}_{0z}}<0$, $s=-1$), the boundary conditions have the form 
\begin{eqnarray}\begin{split}
{{\hat{A}}_{\text{t}1}}\exp (iq_{\text{t}z}^{(1)}l)+{{\hat{A}}_{\text{t}2}}\exp (iq_{\text{t}z}^{(2)}l)&={{\hat{A}}_{\text{in}}}(\Omega),
\\
{{R}_{1}}{{\hat{A}}_{\text{t}1}}+{{R}_{2}}{{\hat{A}}_{\text{t}2}}&=0.
\end{split}
\end{eqnarray}
Thus $(j=1,2)$:
\begin{eqnarray}
\label{B16}
{{\hat{A}}_{\text{t}j}}=\frac{(-P)^{j-1} }{1-P}\hat{A}_{\text{in}}(\Omega )e^{-iq_{\text{t}z}^{(j)}l}.
\end{eqnarray}

For wavenumbers deviating from the central one, i.e., $k={{k}_{0}}+\Omega /c$, while the $x$-component is defined by that of the incident pulse, ${{k}_{x}}=k\sin \theta $, the parameter $\alpha $ is given by
\begin{equation}
\label{B17}
\alpha =\frac{b}{k_{0}^{2}}\left[ 2\left({{k}_{0}}+\frac{\Omega}{c}\right)\sqrt{{{\varepsilon }_{0}}-{{\sin }^{2}}\theta }-b \right] .                
\end{equation}       
Taking into account equations (\ref{B15a}) and (\ref{B16}),  one obtains (\ref{B19}), (\ref{B20}). 
 
\section{Pulse propagation under dynamical Bragg diffraction in the Laue geometry}
\label{app:Laue}
  
Near the Bragg condition in the Laue scheme $2{{k}_{0}}\sin {{\theta }_{B}}=b$, taking into account the continuity of the tangential wavevectors at the boundary $z=0$, we obtain the system for the amplitudes ${{\hat{A}}_{\text{t,r}}}(K,\Omega )$: 
\begin{eqnarray} \label{C4.a}
\begin{split}
({{\varepsilon }_{0}}{{k}^{2}}-q_{\text{t}x}^{2}-q_{\text{t}z}^{2}){{\hat{A}}_{\text{t}}}+{{\varepsilon }_{-s}}{{k}^{2}}{{\hat{A}}_{\text{r}}}&=0, 
\\
{{\varepsilon }_{s}}{{k}^{2}}{{\hat{A}}_{\text{t}}}+[{{\varepsilon }_{0}}{{k}^{2}}-{{({{q}_{\text{t}x}}-sb)}^{2}}-q_{\text{t}z}^{2}]{{\hat{A}}_{\text{r}}}&=0,         \end{split}
\end{eqnarray}
[which is Eqs.~(\ref{dispersion}) for the Laue geometry], where ${{\varepsilon }_{\pm 1}}$ is defined in (\ref{epsilon2}). From the solvability condition of (\ref{C4.a}) we obtain two modes with different $z$-components, $q_{\text{t}z}^{(1,2)}$: 
\begin{multline}
\label{C5}
\left(q_{\text{t}z}^{(1,2)}\right)^{2}={{\varepsilon }_{0}}{{k}^{2}}-k_{\text{t}x}^{2}+\frac{\tilde{\alpha }{{k}^{2}}}{2}\mp
\\
\mp\frac{{{k}^{2}}}{2}\sqrt{{{{\tilde{\alpha }}}^{2}}+{{{{\varepsilon }'}}^{2}}-{{{\tilde{\varepsilon }}}^{2}}(\omega )},    
\end{multline}
where $\tilde{\alpha }=b(2s{{q}_{\text{t}x}}-b)/{{k}^{2}}$ defines the degree of detuning from the exact Bragg condition ${{q}_{\text{t}x}}=sb/2$. The relations connecting the amplitudes of the transmitted, ${\hat{A}_{\text{t}j}}$, and diffracted, ${\hat{A}_{\text{r}j}}$, waves are obtained from Eqs. (\ref{C4.a}): ${{\hat{A}}_{\text{r}j}} = {{R}_{j}}{{\hat{A}}_{\text{t}j}}$, where
\begin{equation}
\label{C7}
{{R}_{1,2}}=\left[\tilde{\alpha }\mp \sqrt{{{{\tilde{\alpha }}}^{2}}+{{{{\varepsilon }'}}^{2}}-{{{\tilde{\varepsilon }}}^{2}}(\omega )}\right]/2{{\varepsilon }_{-s}}.                    
\end{equation}
Now, the Fourier amplitudes of the field ${{\hat{A}}_{\text{t}\text{,r}}}(K,\Omega )$ can be computed from the conditions for the electric and magnetic fields being continuous on the PhC input surface $z=0$: 
\begin{eqnarray}\begin{split}
\label{C8}
{{\hat{A}}_{\text{in}}}+{{\hat{A}}_{S}}&={{\hat{A}}_{\text{t}1}}+{{\hat{A}}_{\text{t}2}}, 
 \\
{{k}_{z}}({{\hat{A}}_{\text{in}}}-{{\hat{A}}_{S}})&=q_{\text{t}z}^{(1)}{{\hat{A}}_{\text{t}1}}+q_{\text{t}z}^{(2)}{{\hat{A}}_{\text{t}2}},   
\\
{{R}_{1}}{{\hat{A}}_{\text{t}1}}&+{{R}_{2}}{{\hat{A}}_{\text{t}2}}=0, \end{split}
\end{eqnarray}
\\
where ${{A}_{S}}$ is the amplitude of the Freshnel reflection, and ${{k}_{z}}=\sqrt{{{(\omega /c)}^{2}}-k_{x}^{2}}$.  From these system we obtain (\ref{C9}).

\end{document}